\documentclass[reprint,aip,apl,groupedaddress]{revtex4-1} 
\usepackage{graphicx}
\usepackage{amsmath}
\usepackage{amssymb}
\usepackage{epsfig}
\usepackage{subeqnarray}
\usepackage{color}
\usepackage{bbold}
\usepackage{dsfont}

\newcommand{\comment}[1]{}

\begin{document}

\title{An alternative approach to efficient simulation of micro/nanoscale phonon transport}

\author{Jean-Philippe M. P\'eraud}
\author{Nicolas G. Hadjiconstantinou}
\affiliation{Department of Mechanical Engineering, Massachusetts Institute of Technology\\ Cambridge, MA  02139, USA}

\date{\today}

\begin{abstract}
Starting from the recently proposed energy-based deviational 
formulation for solving the Boltzmann equation 
[J.-P. P\'eraud and N. G. Hadjiconstantinou, Phys. Rev. B 84,  2011], which  
provides significant computational speedup compared 
to standard Monte Carlo methods for small deviations 
from equilibrium, we show that additional computational 
benefits are possible in the limit that 
the governing equation can be linearized. 
The proposed method exploits the observation that under 
linearized conditions (small temperature differences)  the trajectories 
of individual deviational particles can be decoupled and thus 
simulated independently; this leads to a particularly simple and efficient 
algorithm for simulating steady and transient problems in arbitrary 
three-dimensional geometries, 
{\it without introducing any additional approximation}. 
\end{abstract}

\maketitle


In a previous paper \cite{peraud11}, we presented a low variance Monte Carlo method for solving the Boltzmann transport equation (BTE) for phonons in the relaxation-time approximation whereby computational particles simulate only the deviation from an equilibrium distribution. The benefits of such control-variate formulations \cite{baker05}, which we will refer to as deviational, are twofold: first, in the limit of small temperature differences, deviational methods exhibit substantial computational speedup compared to traditional Monte Carlo methods\cite{peraud11}; this speedup increases quadratically as the characteristic temperature difference goes to zero. Second, by simulating only the deviation from equilibrium,  deviational methods seamlessly and automatically focus the computational effort on regions where it is needed and can thus be used for solving otherwise intractable multiscale problems. In the present article, we show that for problems exhibiting sufficiently small temperature differences such that the BTE can be linearized, deviational computational particles may be treated independently, thus lending themselves to a simulation algorithm that is simpler,  does not use any approximation in space or time, and, depending on the application of interest, can be several orders of magnitude faster than the one presented in Ref.~\onlinecite{peraud11}. 

The deviational approach can be introduced by writing the governing equation (with no approximation) in the form 
\begin{equation}
\frac{\partial e^d}{\partial t} + \mathbf{V}_g \cdot \nabla e^d = \frac{(e^{loc}-e^{eq}_{T_{eq}})-e^d}{\tau}
\label{eq:deviational_BTE}
\end{equation}
where $e^d=e-e^{eq}_{T_{eq}}=\hbar\omega (f-f^{eq}_{T_{eq}}) $, $\tau=\tau(\omega,p,T)$ is the relaxation time ($\omega$, p and T respectively referring to the angular frequency, the polarization and the temperature), $f=f(\mathbf{x},\omega,p,\theta,\phi)$ is the occupation number of phonon states, $\mathbf{V}_g $ is the phonon-bundle group velocity and $f^{eq}_{T_{eq}}=[\exp \left( \hbar \omega/k_b T_{eq} \right)-1]^{-1}$ is a Bose-Einstein distribution at the ``control'' temperature $T_{eq}$ ($k_b$ denotes Boltzmann's constant).
In Ref.~\onlinecite{peraud11} we showed that variance reduction is achieved by simulating only the distribution $De^d$ ($D=D(\omega,p)$ is the density of states) using deviational particles and adding the result due to $De^{eq}_{T_{eq}}$ analytically. According to (\ref{eq:deviational_BTE}), the scattering process is implemented by removing deviational particles from the distribution $D e^d$ (i.e. the current deviational population)  at a rate $\tau(\omega,p,T)^{-1}$ and replacing them with particles drawn from the distribution $D (e^{loc}-e^{eq}_{T_{eq}})/\tau(\omega,p,T)$. 

For small temperature differences, the collision operator in (\ref{eq:deviational_BTE}) can be linearized by writing 
$e^{loc}-e_{T_{eq}}^{eq}\approx(T_{loc}-T_{eq})de^{eq}_{T_{eq}}/dT$
where $T_{loc}$ denotes the local pseudotemperature \cite{peraud11,hao09}. Therefore, scattered particles can be drawn \cite{peraud11} from the distribution 
\begin{equation}
(T_{loc}-T_{eq}) \frac{D(\omega,p)}{\tau(\omega,p,T_{eq})} \frac{de^{eq}_{T_{eq}}}{dT}
\label{gain}
\end{equation}
Since this distribution does not depend on $(T_{loc}-T_{eq}) $ {\it once normalized}, a particle undergoing a scattering event can be drawn from (the normalized form of) (\ref{gain}) without knowledge  of $T_{loc}$; energy conservation is simply ensured by conserving the particle. 
Although this formulation was originally introduced  \cite{peraud11} as a means of truncating the discretization of a semi-infinite simulation domain (by limiting the region where computational cells were used), here, we show that this formulation can be used throughout the computational domain with considerable computational benefits. By removing the need for sampling $T_{loc}$ before processing phonon scattering, the integration timestep  and computational cells found in standard Monte Carlo approaches \cite{mazumder01} are, in fact, unnecessary. Instead, the algorithm proceeds by 
simulating each particle independently and is therefore significantly simpler, requires no discretization in space and time--thereby avoiding the associated numerical error--requires significantly less storage and, depending on the problem of interest, can be several orders of magnitude more computationally efficient.

The proposed algorithm for simulating a particle trajectory between  $t=0$ and $t=t_{final}$ is as follows: 
\begin{itemize}
\item[I] Draw the initial properties (sign $s$, position $\mathbf{x}_0$, frequency $\mathbf{\omega}_0$, polarization $p_0$, direction $\mathbf{\Omega}_0$, and the resulting group velocity vector $\mathbf{V}_{g,0}$) of the particle. For time-dependent calculations, also set up the initial time $t_0$ of the particle (see below). 
\item[II] Calculate the traveling time until the first scattering (relaxation) event: uniformly draw a random number $R\in]0,1[$ and calculate $\Delta t=-\tau(\omega_0,p_0,T_{eq}) \ln(R) \label{eq:time}$

\item[III] Calculate $\tilde{\mathbf{x}}_{new}=\mathbf{x}_{0}+\mathbf{V}_{g,0} \Delta t$. Search for collisions with system boundaries in the time interval $\Delta t$.
\item[IVa] If a collision with a system boundary occurs, say at $\mathbf{x}_b$, set $\mathbf{x}_{new}=\mathbf{x}_b$ and 
update the internal time $t_{new}=t_0+||(\mathbf{x}_b-\mathbf{x}_{0})||/||\mathbf{V}_{g,0}||$. Depending on the nature of the reflection (specular or diffuse), set the new traveling direction appropriately (as explained for example in Ref.~\onlinecite{mazumder01}).
\item[IVb] If no collision with system boundaries occurs, the particle undergoes scattering at position $\mathbf{x}_{new}=\tilde{\mathbf{x}}_{new}$. The internal time is updated to $t_{new}=t_0+\Delta t$. New frequency $\omega_{new}$ and polarization $p_{new}$ are then drawn from (\ref{gain}). A new traveling direction is also chosen: in this work, we consider isotropic scattering, but this can easily be generalized to non-isotropic scattering. From these parameters, a new velocity vector $\mathbf{V}_{g,new}$ can be defined. The particle sign remains unchanged by scattering.
\item[V] Sample the contribution of segment [$\mathbf{x}_0$,$\mathbf{x}_{new}$] to macroscopic properties (see below).
\item[VI] If $t_{new}>t_{final}$, proceed to step I to begin simulation of the next particle; otherwise, set $\{ .\}_{0}=\{ . \}_{new}$, where $\{.\}$ denotes the set of all properties of the particle, and return to step II. 
\end{itemize}

The total number of particles processed, $N$, is determined by the total amount of deviational energy involved in the phenomenon of interest divided by the effective energy  carried by each computational particle, $\mathcal{E}_{eff}$. The latter is  chosen such that the resulting number of computational particles balances computational cost with the need for low statistical uncertainty. The contribution of initial and boundary conditions to the deviational population can be treated by specialized source terms. Denoting  the {\it sum} of all source terms (including boundary and initial conditions) by $Q(\mathbf{x},\omega,\Omega,p,t)$, each particle's initial time $t_0$ is randomly drawn by inverting the generalized cumulative distribution $\int_{t^\prime=0}^t \sum_p \int\int\int Q d\mathbf{x}d\omega d\mathbf{\Omega}dt^\prime$. For example, in a finite 1D system parametrized by the space coordinate $x$, the contribution of the initial condition (say initial temperature $T_{i}(x)$ at $t=0$) to $Q$ is $(4\pi)^{-1}D |e^d_i|\delta(t)=C_{\omega,p}|T_i(x)-T_{eq}|\delta(t)$, where $C_{\omega,p}=(4\pi)^{-1}Dde^{eq}_{T_{eq}}/dT$; the contribution of an isothermal boundary at $x=0$ and at temperature $T_{b}(t)$ to the half space $x>0$ is $(4\pi)^{-1}D |e^d_b| \mathbf{V}_{g} \cdot \mathbf{\hat{e}}_x\delta(x)H[\mathbf{V}_{g} \cdot \mathbf{\hat{e}}_x]=V_g \cos(\theta) C_{\omega,p}|T_b(t)-T_{eq}|\delta(x) H[\cos(\theta)]$, where $\theta$ is the angle with respect to the $x>0$ direction, and H the Heaviside step function.

We now discuss the sampling process in more detail. \\Let $I_g(t^\prime)=\sum_{p} \int \int \int (4 \pi)^{-1}D g e^d (t^\prime) d\mathbf{x}d\omega d\mathbf{\Omega}$ be the macroscopic property of interest (at time $t^\prime$) in terms of a general microscopic property $g=g(\mathbf{x},\omega,p,\mathbf{\Omega})$. Recalling that the deviational simulation approximates the distribution $e^d$ in phase space using deviational (computational) particles \cite{peraud11}, the estimate of $I_g(t^\prime)$ is given by
\begin{equation}
\tilde{I}_g(t^\prime)=\mathcal{E}_{eff} \sum_{i}  s_i  g\left[\mathbf{x}_i(t^\prime),\omega_i(t^\prime),p_i(t^\prime),\mathbf{\Omega}_i(t^\prime)\right]
\label{Ig_t}
\end{equation}
where symbols have their usual meanings and $s_i$ is the sign of deviational particle $i$.
For example, if the quantity of interest is the $z$-component of the heat flux vector in some region of space $\mathbf{R}$ with volume $\mu(\mathbf{R})$ and defined by the characteristic function $\chi_\mathbf{R}$, then $g=\mathbf{V}_{g}\cdot \hat{\mathbf{e}}_z \chi_\mathbf{R}/\mu(\mathbf{R})$ and thus particle $i$ only contributes to $\tilde{I}_g(t^\prime)$ if $\mathbf{x}_i(t^\prime)$  [its position at $t^\prime$---calculated by linear interpolation between ($\mathbf{x}_0$,$t_0$) and ($\mathbf{x}_{new}$,$t_{new}$)] is in $\mathbf{R}$.

As in standard Monte Carlo methods, steady problems can be sampled
by replacing ensemble-averaging with time-averaging
$\bar{I}_g(ss)=(1/{\cal T})\int_{t^\prime=t_{ss}}^{t_{ss}+{\cal T}} \tilde{I}_g(t^\prime)d t^\prime=({\cal E}_{eff}/{\cal T})\sum_{i} \int_{t^\prime=t_{ss}}^{t_{ss}+{\cal T}} s_i g dt^\prime$ over a time
period ${\cal T}$, provided sufficient time $t_{ss}$ has passed for steady
conditions to prevail. Computational benefits can be realized 
by noting that for steady conditions to be possible, the system must be under the influence of only steady particle sources ($Q\neq Q(t)$). 
By taking the limit ${\cal T}\rightarrow +\infty$, the influence of initial conditions vanishes, allowing the simulation to directly solve for---and thus focus all computational effort on---the steady state. Particles are sampled over their complete trajectories, from emission (by the steady sources) to termination (which happens for example through absorption by a boundary), using 
\begin{equation}
\tilde{I}_g=\dot{\mathcal{E}}_{eff}\sum_{i} s_i \int g\left[\mathbf{x}_i (t),\omega_i (t),p_i(t), \mathbf{\Omega}_i(t)\right] dt
\label{eq:integral_contribution}
\end{equation}
because in this limit the effective deviational power from the steady sources $\dot{{\mathcal E}}_{eff} \equiv {\mathcal E}_{eff}/(t_{ss}+{\cal T})$ reduces to ${\mathcal E}_{eff}/{\cal T}$.

Mathematical proofs of this statement can be found in the linear transport theory literature (see for example Ref.~\onlinecite{spanier69}). In the case of the heat flux in the $z$-direction ($g=V_z \chi_{\mathbf{R}}/\mu(\mathbf{R})$) averaged over the domain $\mathbf{R}$ discussed above, equation (\ref{eq:integral_contribution}) reduces to $\dot{\mathcal{E}}_{eff} \sum_{i}  s_i {\cal L}_i/\mu(\mathbf{R})$ where ${\cal L}_i$ is the \textit{total} algebraic length  traveled in the $z$-direction by particle $i$ while in $\mathbf{R}$ (can be negative if traveling in negative direction).

\begin{figure}[htbp]
	\centering
	\includegraphics[width=.45\textwidth]{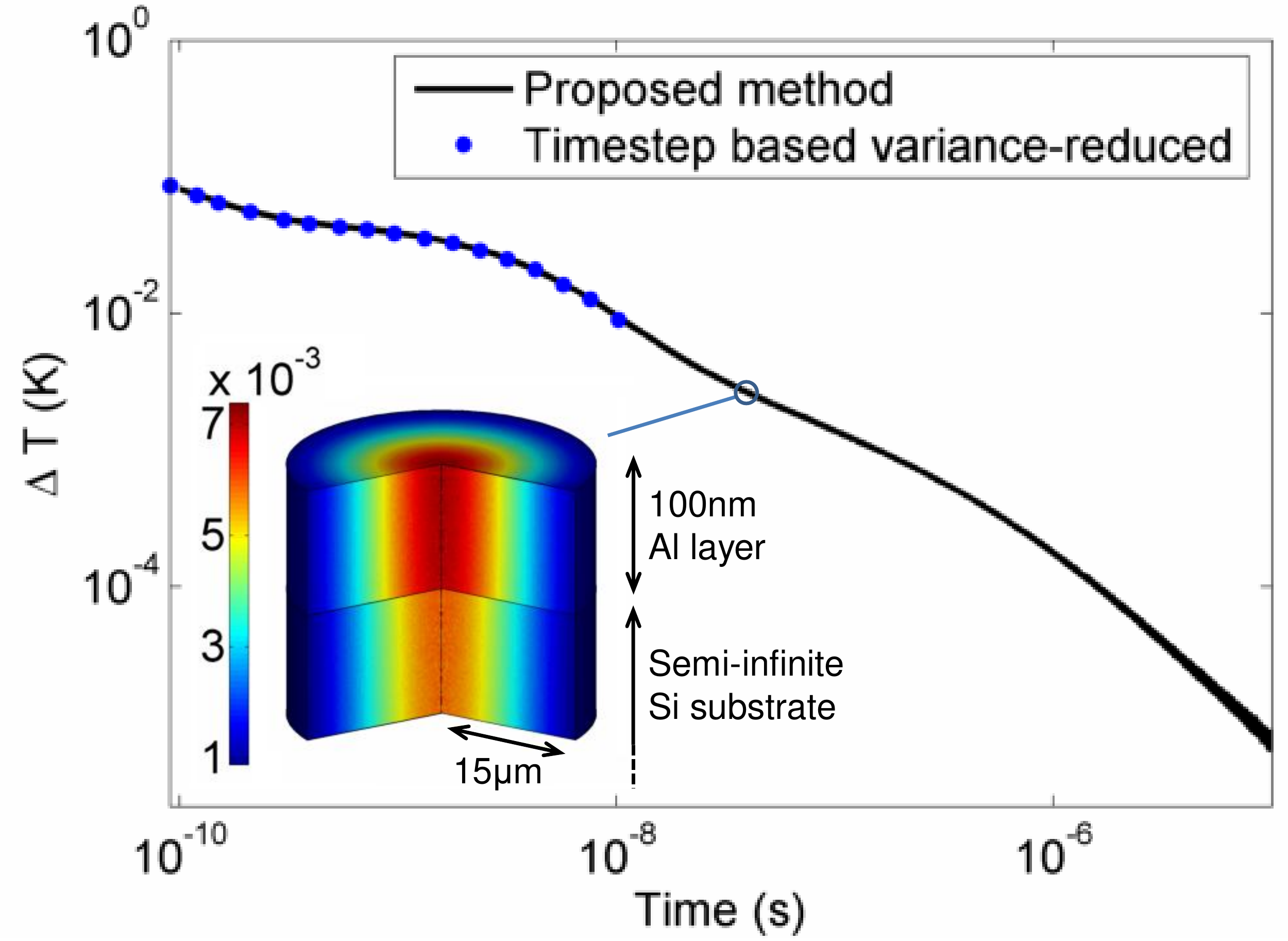}
	\caption{Surface temperature (calculated as the spatial average in a cylinder of radius 10 $\mu$m and height 5nm) in the TTR experiment as a function of time, calculated with the variance-reduced Monte Carlo method using timesteps (see Ref.~\onlinecite{peraud11}), and with the proposed method. The latter reaches significantly longer times.}
	\label{fig:TTR_long_times}
\end{figure}

The proposed algorithm has been extensively validated using a number of test problems \cite{peraud12} including the thin film problem described in Ref.~\onlinecite{peraud11} for which an analytical solution exists. Here, we present simulation results from two problems of practical interest (we use the same materials and phonon properties as in Ref.~\onlinecite{peraud11} and \onlinecite{peraud12}). 
First, we consider the transient thermo-reflectance (TTR) experiment presented in Ref.~\onlinecite{schmidt08}  and used in Ref.~\onlinecite{minnich11} as a thermal conductivity spectroscopy technique. Using the algorithm described above, we simulate the thermal response of a thin film of aluminum on a substrate of silicon  after a laser pulse irradiates the surface and provides localized heating at $t=0$. More details on the problem formulation can be found in Ref.~\onlinecite{peraud11}, where it is also shown that the deviational formulation enabled the simulation of the temperature field in this three-dimensional problem for several nanoseconds (due to the small temperature differences involved, simulation using standard Monte Carlo methods is too expensive). The {\it additional} speedup due to the  present algorithm allows us to calculate the response to a single pulse up to 10 $\mu$s (Fig. \ref{fig:TTR_long_times}). Ultimately, we expect this improvement to be invaluable towards the computational description of the phonon spectroscopy experiment discussed in Ref.~\onlinecite{minnich11,minnich11bis}.

\begin{figure}[htbp]
	\centering
	\includegraphics[width=.45\textwidth]{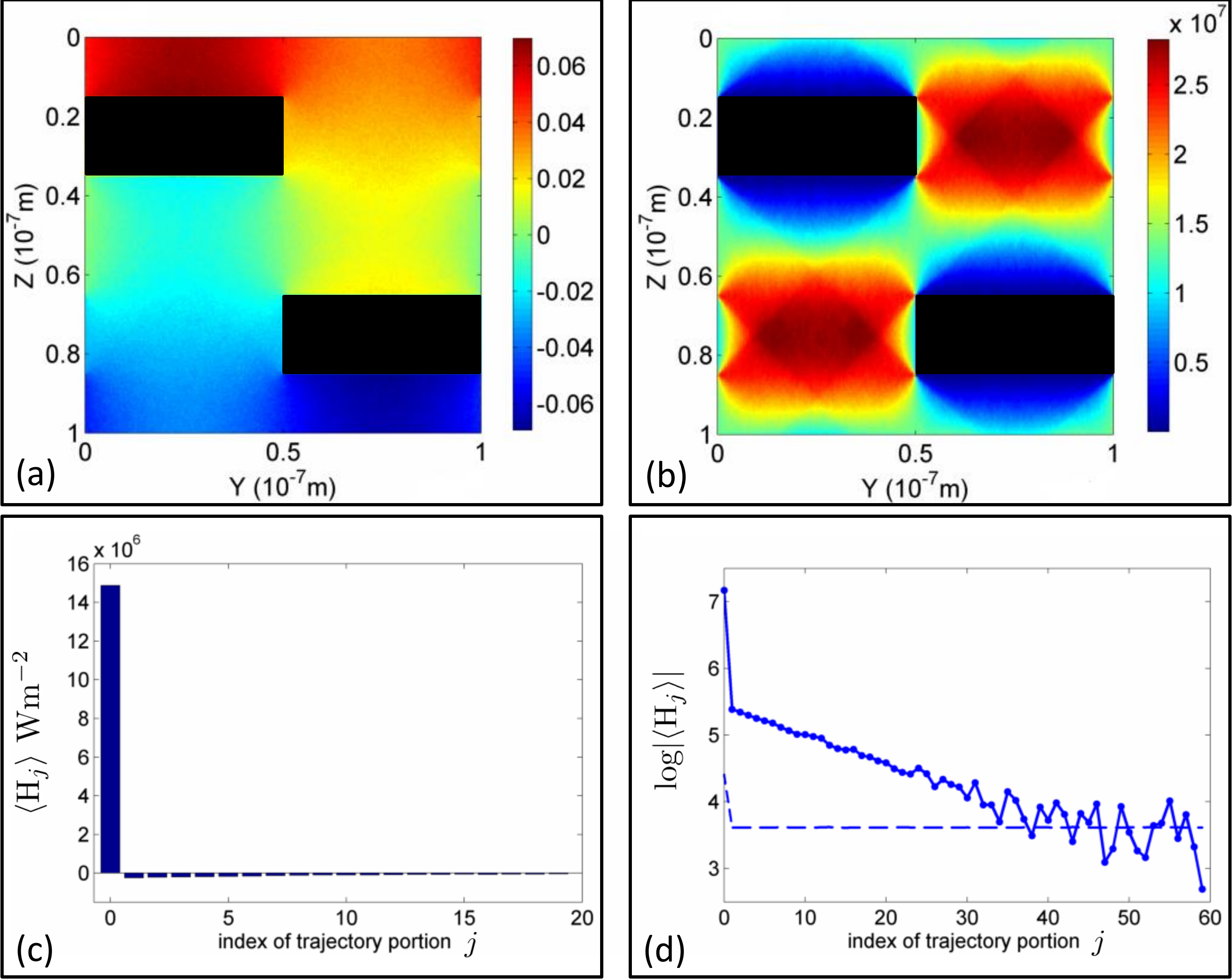}
	\caption{(a) Local temperature field ($T-T_0$) expressed in kelvin in a periodic nanostructure subject to a temperature gradient of $-10^6$Km$^{-1}\mathbf{\hat{e}}_z$. (b) Local heat flux (Wm$^{-2}$) in the $z-$direction. (c) Average particle contribution to the heat flux as a function of the particle's scattering event number. On average, contributions after the first scattering event amount to approximately 20\% of the total heat flux. (d) Comparison between the absolute value in the heat flux contributions and their associated statistical uncertainty $\sigma/\sqrt{N}$ (dashed line); $\sigma$ is the standard deviation in the heat flux as measured from simulation data.}
	\label{fig:heat_flux}
\end{figure}

As a second application, we consider the determination of the thermal conductivity of complex periodic nanostructures, which has recently received a lot of attention in the literature \cite{hao09, chen97, huang09}. Here we consider a periodic nanostructure with a unit cell as shown in Fig. \ref{fig:heat_flux} in the presence of a temperature gradient in the $z$-direction. By calculating the heat flux in the direction of the gradient, we can determine the ``effective'' thermal conductivity of the nanostructure.
Instead of considering an equilibrium $T_{eq}$ that is spatially constant, we allow the latter to vary in space. This approach has been shown to improve variance reduction \cite{radtke09} because it allows the control temperature to follow the physical temperature more closely; it is particularly convenient for imposing external fields such the one considered here, in which  $T_{eq}(\mathbf{x})$ varies linearly from $T_1$ to $T_2>T_1$. With this choice of $T_{eq}$, the BTE becomes
\begin{equation}
\frac{\partial e^d}{\partial t} + \mathbf{V}_g \cdot \nabla e^d = \frac{(e^{loc}-e^{eq}_{T_{eq}(\mathbf{x})})-e^d}{\tau} -  \mathbf{V}_g \cdot \nabla e^{eq}_{T_{eq}(\mathbf{x})}
\label{eq:driving_force}
\end{equation}
where the last term on the right hand side can be interpreted as a {\it volumetric} 
source of deviational particles due to the imposed temperature gradient.
When simulating (\ref{eq:driving_force}), the periodic nature of the calculation is straightforwardly implemented 
by requiring that positive and negative (deviational) particles individually obey periodic boundary conditions. 
Note that since the BTE is not linearized in (\ref{eq:driving_force}), the source term formulation is valid for all deviational methods (e.g. Ref.~\onlinecite{peraud11}). 

In order to avoid non-linearities in the response, and because our simulation method does not require large temperature differences for accuracy, we will assume small temperature differences $(T_2-T_1)/T_0\ll 1$, with $T_0=(T_1+T_2)/2$; material properties, such as $\tau(\omega,p,T)$, as well as the distribution $de^{eq}_{T_{eq}}/dT$ will be evaluated at $T_0$. In other words, in the linear regime, the source term in (\ref{eq:driving_force}) is uniform in space.

The simulation proceeds as outlined above (steady state sampling), with a few additional features due to the periodicity of the problem. Particles are drawn from
\begin{equation}
-\frac{D(\omega,p)}{4\pi}V_g(\omega,p) \cos(\theta) \frac{de^{eq}_{T_{0}}}{dT}\frac{dT}{dz}
\label{eq:emission}
\end{equation}
where $\theta$ is the polar angle (measured with respect to the $z$ axis) of the particle traveling direction. Due to symmetry, the same number of negative and positive particles should be emitted. Particles exiting the domain are periodically reinserted.

The absence of absorbing boundaries coupled to energy (particle) conservation results in infinitely long particle trajectories which 
are impossible to track numerically. To overcome this, we use the observation that after several scattering events,  
particle properties are almost completely randomized (i.e. independent from the initial state at emission) and thus can be  
terminated with only a small effect on the simulation accuracy.   
Fig. \ref{fig:heat_flux} illustrates  this for the case of the 2D nanostructure 
presented in the same figure: the mean heat flux contribution (averaged over many  different 
particle trajectories) between scattering event $j$ and $j+1$, denoted by $\langle H_j\rangle$, 
decreases rapidly as $j$ increases. 
Fig. \ref{fig:heat_flux}(d) shows that after approximately 40
scattering events the statistical uncertainty in $\langle H_{j} \rangle$ (calculated in the present case using $N=8 \cdot 10^6$ particles) becomes on the order of $\langle H_{j}\rangle $, suggesting that the benefit from collecting further samples is minimal and 
terminating the particle is justified. Furthermore, the error in the estimate of the heat flux can be controlled thanks to the exponential decay we observe in $\langle H_{j} \rangle$ after a few scattering events. Development of a theoretical prediction for the number of scattering events a particle must undergo before it can be discarded and its dependence on the problem characteristics is the subject of ongoing research work. For the moment, this criterion can be determined empirically as shown here. Figures \ref{fig:heat_flux}(a) and (b) show the result obtained using this approach. The calculated thermal conductivities ($9.8$Wm$^{-1}$K$^{-1}$ in the configuration of Fig. \ref{fig:heat_flux}) are in agreement with previous results \cite{peraud11}, while the computational time was reduced by approximately 2 orders of magnitude.

An interesting special case of the above problem is the calculation of thermal conductivities 
of thin films with diffuse boundaries (parallel to the $z$-direction). 
In this case, using (\ref{eq:driving_force}) and (\ref{eq:emission})
to describe the imposed temperature gradient  
reduces the problem dimensionality to one, namely the direction normal to the diffuse boundaries. The resulting problem 
is sufficiently simple (it admits an analytical solution) 
and in the case of isotropic scattering and diffuse walls is of 
sufficiently high symmetry, that the contribution to the heat flux 
after the first wall collision or the first scattering event vanishes, because the expected value of 
$[\mathbf{V}_{g}]\cdot \hat{\mathbf{e}}_z$ is zero.
This observation can be used to put in context 
 the results presented in Ref.~\onlinecite{mcgaughey12} where  
the thermal conductivity of nanostructures was approximately calculated using a Monte Carlo approach which follows ``test'' phonon paths to 
their first free path termination (due to either a boundary or relaxation).
This treatment yields the correct result for a thin film due 
to the simplicity and symmetric nature of this problem; 
under more general conditions, terminating particle trajectories 
after the first collision event and assuming 
Fourier's law to be valid as assumed in Ref.~\onlinecite{mcgaughey12},
leads to an inaccurate answer. (In the case of the problem shown in Fig. 
\ref{fig:heat_flux}, it leads to a value for the average heat flux/thermal
conductivity that is approximately 250$\%$ larger).

Our theoretical formulation above also provides justification  
for two of the assumptions used in Ref.~\onlinecite{mcgaughey12}, namely that the 
free paths follow a Poisson distribution 
(see equation (\ref{eq:deviational_BTE})) and 
that ``test'' particles are emitted 
from any point of the nanostructure with equal probability. The latter is 
only true because, as stated above, under linearized 
conditions, the source term (\ref{eq:emission}) is constant (and particles can 
be terminated after their first scattering event in the thin-film problem). 
However, as stated 
above, for calculating the thermal conductivity of nanostructures, unless 
symmetry allows, deviational particles need to be tracked well beyond their 
first free-path termination. This is also true for solving the 
Boltzmann equation under general 
conditions, since only then the correct non-equilibrium distribution of 
deviational carriers is obtained \cite{radtke09,baker05,wagner08,jht10}.

We conclude by emphasizing that the only approximation introduced in this work 
comes from the assumption that the governing BTE can be 
linearized. As shown above, this is reasonable for a number of applications of interest. Under this condition, the proposed algorithm is in fact 
``more accurate'' than alternative algorithms since it involves no timestep 
or spatial discretization. We also note that under this formulation deviational particles  
share similarities with neutrons which also do not interact. 
Given the substantial literature on neutron transport simulation \cite{spanier69}, 
the room for improvement and gain in efficiency in the proposed formulation is 
considerable. 

This work was supported in part by the the Singapore-MIT alliance.  J-P.M.P. 
gratefully acknowledges financial support
from the Total MIT Energy Initiative Fellowship.

\bibliographystyle{plainnat}

\end{document}